\documentclass[MSNbibl,nameyear,dvips]{arxstspdf}
\usepackage{flushend}
\usepackage{stfloats}
\usepackage{graphicx}
\usepackage{url,breakurl}
\volume{29}
\issue{2}
\pubyear{2014}
\firstpage{201}
\lastpage{213}
\doi{10.1214/14-STS477} % kopijuoti is 'New paper accepted'
\makeatletter
\renewcommand{\citet}[1]{\citeauthor{#1} (\citeyear{#1})}
\renewcommand{\citep}[1]{(\cite{#1})}
\makeatother

\begin{document}
\begin{frontmatter}

\title{Reactive Programming for Interactive Graphics}%\thanksref{T1}
% kai straipsnis turi susijusiu diskusiju ir rejoinder'iu
%rejoinder at \relateddoi{r}{10.1214/00-STSXXXX}.}
\runtitle{Reactive Programming for Interactive Graphics}

\begin{aug}
\author[a]{\fnms{Yihui}~\snm{Xie}\corref{}\ead[label=e1]{xie@yihui.name}\ead[label=u1,url]{http://yihui.name}},
\author[b]{\fnms{Heike}~\snm{Hofmann}\ead[label=e2]{hofmann@iastate.edu}\ead[label=u2,url]{http://hofmann.public.iastate.edu}}
\and
\author[a]{\fnms{Xiaoyue}~\snm{Cheng}\ead[label=e3]{xycheng@iastate.edu}\ead[label=u3,url]{http://xycheng.public.iastate.edu}}
\runauthor{Y. Xie, H. Hofmann and X. Cheng}

\affiliation{Iowa State University}

\address[a]{Yihui Xie and Xiaoyue Cheng are Ph.D. Students, Department of Statistics,
Iowa State University,
102 Snedecor Hall,
Ames, Iowa 50011,
USA \printead{e1,u1,e3,u3}.}
\address[b]{Heike Hofmann is Professor,
Department of Statistics,
Iowa State University,
2413 Snedecor Hall,
Ames, Iowa 50011,
USA \printead{e2,u2}.}
\end{aug}

% ABSTRACT
%
\begin{abstract}
One of the big challenges of developing interactive statistical applications
is the management of the data pipeline, which controls transformations from
data to plot. The user's interactions needs to be propagated through these
modules and reflected in the output representation at a fast pace. Each
individual module may be easy to develop and manage, but the dependency
structure can be quite challenging. The MVC (Model/View/Controller) pattern
is an attempt to solve the problem by separating the user's interaction from
the representation of the data. In this paper we discuss the paradigm of
\emph{reactive programming} in the framework of the MVC architecture and
show its applicability to interactive graphics. Under this paradigm,
developers benefit from the separation of user interaction from the
graphical representation, which makes it easier for users and
developers to
extend interactive applications. We show the central role of reactive data
objects in an interactive graphics system, implemented as the R package
\textbf{cranvas}, which is freely available on GitHub and the main
developers include the authors of this paper.
\end{abstract}

% KEYWORDS
\begin{keyword}
\kwd{Reactive programming}
\kwd{interactive graphics}
\kwd{R language}
\end{keyword}
\end{frontmatter}

%%==============================================================================%
%s1 #&#
\section{Introduction}
%%==============================================================================%

Interactive graphics progresses us beyond the limitations of static
statistical displays, in particular, for exploring multidimensional data.
With a static image, we can only see one aspect of the data at a time.
Interactive graphics allows us to inspect data dynamically from multiple
views. For example, we may draw a scatterplot of two variables and a stacked
bar chart showing the proportions of missing values for the rest of the
variables
in a data set (two stacked bars per variable). Then we can highlight
the bar
that indicates the missing values of one variable, and the subset of points
corresponding to these missing values in the scatterplot are highlighted
immediately, so we can examine the conditional bivariate relationship
in the
scatterplot.

The term ``interactive graphics'' can be ambiguous, as disclosed by
\citet{swayne1999} in an editorial of \textit{Computational Statistics}: it may
imply the direct manipulation of the graph itself, manipulation of the graph
controls or even the command-line interaction with graphs. We
primarily mean the direct manipulation on graphs, but other meanings still
have their usefulness. For instance, we may change the bin width of a
histogram through a slider or brush all the outliers in a scatterplot using
a command line with a numeric criterion, achieving a higher degree of control
than direct manipulation allows.

The main tasks that an interactive statistical graphics system should
support are as follows:
\begin{longlist}[2.]
\item[1.] Single display interactions, such as modifying the plot attributes
(brushing, zooming, panning, deletion) and obtaining additional
information (querying graphical elements);

\item[2.] Linking between different displays of the same data set or related
data sets. For example, suppose we have a scatterplot of the variable $Y$
versus $X$ and a histogram of $Z$ (all three variables are from the same
data set), when we highlight a subset of points in the scatterplot and we
need to show the distribution of the subset of $Z$ in the histogram as well.
\end{longlist}

The first set of tasks is easier to solve---there are a lot of web
applications that allow various single display interactions, for instance,
Gapminder \citep{RJ2009}, ManyEyes \citep{viegas2007}, JMP \citep
{jmp} and
D3 \citep{bostock2011d3}.

Linked graphics, which is less common, allows changes across different
displays as well as across different aggregation levels of the data.
The key
difficulty is how to let the plots be aware of each other's changes and
respond both automatically and immediately. There are several types of
linking between plots, one-to-one linking, categorical linking and
geographical linking \citep{dykes1998}; see \citet{hurley1988},
\citet{stuetzle1987} and \citet{mcdonald1990} for some early demonstrations and
implementations. We will show how linking is related to, and achieved by,
reactive programming in this paper.

%f1 #&#
\begin{figure*}

\includegraphics{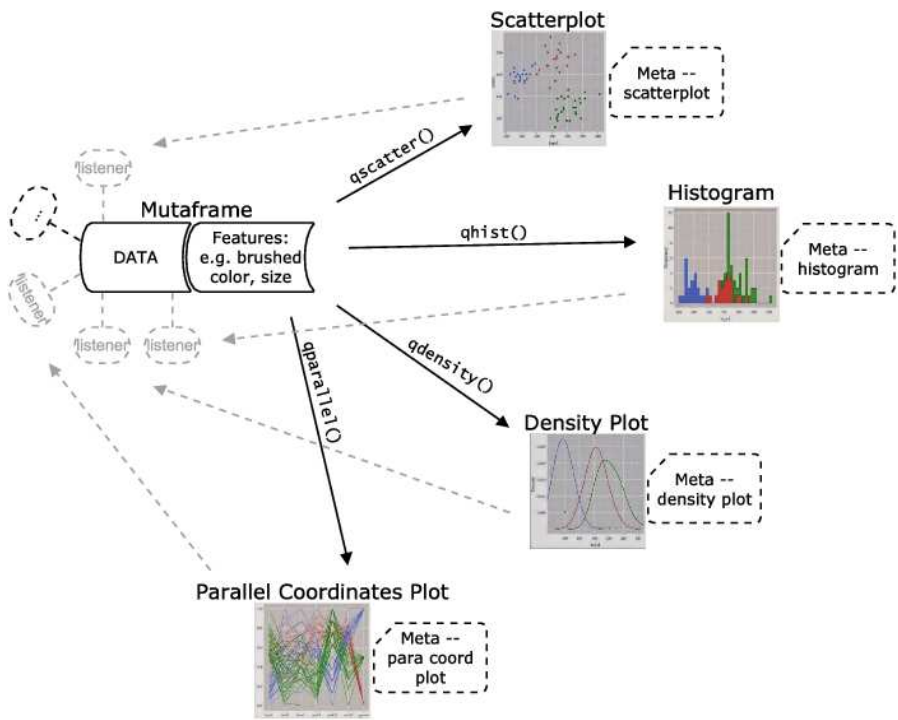}

  %[A representation of the pipeline in the \textbf{cranvas} package]
  \caption{A representation of the pipeline in the \textbf{cranvas}
  package. It shows
how plots are linked to each other as well as to the data source.
Elements are
added to the data to facilitate the interaction. Arrows with solid lines
indicate R functions that draw the plots. Dashed lines indicate reactions
fed back to the data elements.}\label{fig:pipeline}\vspace*{6pt}
\end{figure*}

A number of stand-alone systems for interactive \mbox{statistical} graphics exist.
Early systems include PRIM-9, an interactive computer graphics system to
picture, rotate, isolate and mask data in up to 9 dimensions \citep{prim9}.
Data Desk \citep{velleman1988} and LISP-STAT \citep{tierney1990} provided
tight integration with interactive graphics as well as numerical modeling.
In particular, LISP-STAT is also a programmable environment like R
\citep{R2013}, but, unfortunately, today R is the more popular choice.
\citet{tierney2005} described a few desirable approaches toward programmable
interactive graphics, which were not implemented in LISP-STAT due to
limitations of the toolkit. These are all relatively straightforward in the
framework of R and Qt \citep{Qt}. XGobi and GGobi (\cite{swayne2003}; \cite{cook07}),
MANET \citep{unwin96} and Mondrian \citep{theus02} support
interactive displays of multivariate data, but lack extensibility and a
tight integration with modeling in R. The \textbf{rggobi} package
\citep{rggobi} is an interface between R and GGobi based on the GTK+ toolkit. The
\textbf{iplots} package \citep{R-iplots} provides high interaction
statistical graphics; it is written in Java using the Swing toolkit and
communicates with R through the \textbf{rJava} package.

% TODO: justify why cranvas instead of iplots

One of the big challenges in the development of interactive statistical
applications is to resolve a user's action on the data level. This is
sometimes referred to as the ``plumbing'' of interactive graphics.
\citeauthor{buja1988}
(\citeyear{buja1988}, page 298) introduced the concept of a viewing pipeline for data
plots. The pipeline takes the raw data, through transformation,
standardization, randomization, projection, viewporting and graphical element
in a plot. Some components of the pipeline can be made implicit, such
as the
so-called ``window-to-viewport'' transformation (i.e., viewporting),
due to
technological advances in computer graphics toolkits. For example, Qt can
take care of such low-level details automatically.
\citet{wickham2009plumbing} outlined a more general pipeline for interactive
graphics, but it did not cover implementation details, which is the
focus of
this paper.

The R package \textbf{cranvas} \citep{cranvas} is an interactive graphics
system built under the classical Model/View/Controller (MVC) architecture
and adopts the reactive programming paradigm to achieve interactivity.
Figure~\ref{fig:pipeline} shows a basic pipeline in the \textbf{cranvas}
package and its most important components, mutaframes and metadata objects,
which are ``reactive'' by design. The pipeline starts with a data
source (a
mutaframe) as the central commander of the system. Any plot can modify the
data source as the user interacts with the plot and, as soon as the
mutaframe is modified, its reactive nature will propagate the changes
to all
other plots in the system automatically. In addition, each plot also
has its
own attributes that are described by the metadata beyond the
mutaframes. A
metadata object is also reactive, but it is only linked to a specific plot.
For example, the bin width of a histogram is stored in its metadata, and
when the user's action induces a change in this value, the histogram
responds accordingly.

The paper is organized as follows. We start with a discussion of the MVC
design. Section~\ref{sec:reactive} describes the reactive programming
paradigm relative to the MVC architecture, using \textbf{cranvas} as an
example. Section~\ref{sec:interaction} provides specific examples of how
interaction is realized.

%%==============================================================================%
%s2 #&#
\section{The MVC Architecture}\label{sec:mvc}
%%==============================================================================%

MVC is a software architecture design described originally by Trygve
Reenskaug in
the 1970s and in detail by \citet{Krasner1988}. It is widely
used in GUI (Graphical User Interface) applications, including web
applications \citep{leff2001web}. There have been a number of R packages
utilizing the MVC architecture. For example, \citet{Whalen05} built an
interactive GUI system using the MVC design to explore linked data. The GUI
was based on the \textbf{RGtk} package, which later evolved into
\textbf{RGtk2}
\citep{RGtk2}, and MVC was implemented in the \textbf{MVCClass} package.

The main reason for the popularity of MVC is because it minimizes the
dependencies between different components of an application. For example,
let us assume that the model component consists of a data transformation,
such as a square-root or log transformation. The model does not depend on
the view, but the view depends on the model in the sense that if the data
is changed or a different data transformation is chosen, the view has
to be
updated to reflect this change. The model developer therefore never
needs to
deal with the representation of the data on the screen.

In a traditional MVC design, the controller sends commands to both the model
and the view to update their states. Below is a minimal example in R
code on
how to brush a scatterplot under the MVC design.

%f2 #&#
\begin{figure}[h]

\includegraphics{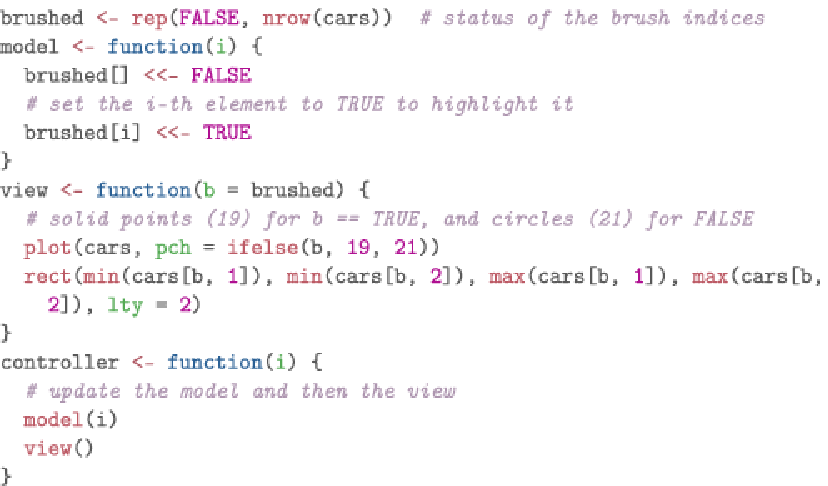}

  %\begin{knitrout}
%{fgcolor} \ifinner\ifhmode \def\at@end@of@kframe{\end{minipage}}
%
%-\width
%%
%{,} \hlkwd{nrow}\hlstd{(cars))} \hlcom{# status of the brush indices}
%{) \{}
%{= brushed) \{}
%{ifelse}\hlstd{(b,} \hlnum{19}\hlstd{,} \hlnum{21}\hlstd{))}
%{]),} \hlkwd{min}\hlstd{(cars[b,} \hlnum{2}\hlstd{]),} \hlkwd
%{max}\hlstd{(cars[b,} \hlnum{1}\hlstd{]),} \hlkwd{max}\hlstd{(cars[b,}
%{i}\hlstd{) \{}
%%
 % \caption{}
 \label{fig2}
\end{figure}

When the user brushes the scatterplot, we can obtain the indices of the
points under the brush rectangle (denoted by \texttt{i} in the code above).
Then we pass the indices to the model to change
the brush status (the vector
\texttt{brushed}) and redraw the plot. %See Figure~\ref{fig3}.

%f3 #&#
\begin{figure}[h]

\includegraphics{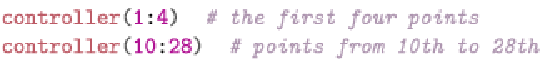}

%{fgcolor} \ifinner\ifhmode \def\at@end@of@kframe{\end{minipage}}
%%
%-\width
%%
%%
%
  %\caption{}
  \label{fig3}
\end{figure}

Decoupling the system into three components enables components to be
accessed independently. For example, we can call the model or the view
separately without modifying their source code.

The problem with the traditional MVC design is that we have to be explicit
about updating the model and the view in the controller. In the context of
interactive graphics, this can be a burden for developers. For instance,
when there are multiple views in the system, the controller must notify all
views explicitly of all of the changes in the system. When a new view is
added to the system, the controller must be updated accordingly. Below
is
what we normally do when we add a new view to the system.

%f4 #&#
\begin{figure}[h]

\includegraphics{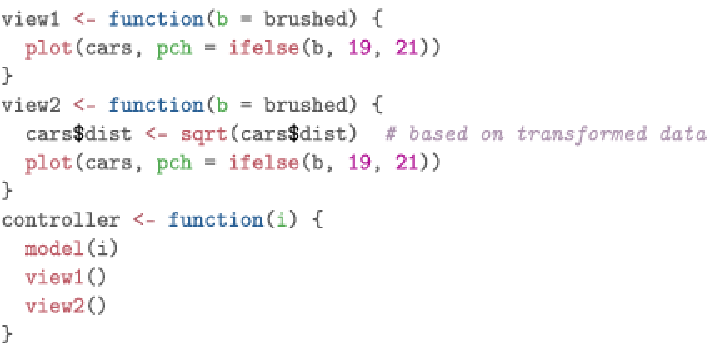}

%{fgcolor} \ifinner\ifhmode \def\at@end@of@kframe{\end{minipage}}
%%
%-\width
%%
%{= brushed) \{}
%{ifelse}\hlstd{(b,} \hlnum{19}\hlstd{,} \hlnum{21}\hlstd{))}
%{= brushed) \{}
%{(cars}\hlopt{$}\hlstd{dist)} \hlcom{# based on transformed data}
%{ifelse}\hlstd{(b,} \hlnum{19}\hlstd{,} \hlnum{21}\hlstd{))}
%{i}\hlstd{) \{}
%%
  %\caption{}
  \label{fig4}
\end{figure}

%%==============================================================================%
%s3 #&#
\section{Reactive Programming}\label{sec:reactive}
%%==============================================================================%

Reactive programming is an object-oriented programming paradigm based
on an
event-listener model and targeted at the propagation of changes in data
flows. We attach listeners on data objects such that (different) events will
be triggered corresponding to changes in data. In the above example, the
plot will be updated as soon as the object \texttt{brushed} is modified
without the need to explicitly call \texttt{view()}. This makes it much
easier to express the logic of interactive graphics. We will discuss
how it
works and its application in \textbf{cranvas}.
Shiny \citep{shiny} is another application of reactive programming in
the R
community which makes it easy to interact between HTML elements and R, but
it does not have a specific emphasis on statistical graphics.

To provide interactive graphics in \textbf{cranvas}, there are two
types of
objects:
\begin{itemize}

\item data presented in the plots, often of a tabular form like data
frames in R

\item metadata to store additional information of the plots such as the
axis limits; it is irregular like a list in R.
\end{itemize}

There are two approaches for making objects reactive:
\emph{mutaframes} \citep{R-plumbr} for the data object and
\emph{reference classes} \citep{Chambers2013} for the metadata.
The fundamental technique underlying them
is the \emph{active binding} in R, thanks to the work of the R Core
team\vadjust{\goodbreak} (in
particular, Luke Tierney). For details, see the documentation of
\texttt
{makeActiveBinding} in R. Both mutaframes and reference classes use active
bindings to make elements inside them (such as data columns or list
members) reactive whenever they are modified.

Active bindings allow events (expressed as functions) to be attached on
objects and these events are executed when objects are assigned new values.
Below is an implementation with active bindings, expanding on the example
code in the previous section.

%f5 #&#
\begin{figure}[h]

\includegraphics{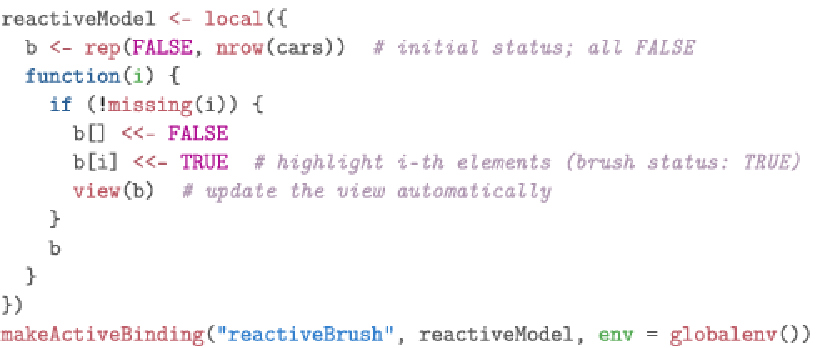}

%{fgcolor} \ifinner\ifhmode \def\at@end@of@kframe{\end{minipage}}
%%
%-\width
%%
%elements (brush status: TRUE)}
%reactiveModel,} \hlkwc{env} \hlstd{=} \hlkwd{globalenv}\hlstd{())}
%%
  %\caption{}
  \label{fig5}
\end{figure}

\begin{figure}\vspace*{-6pt}

\includegraphics{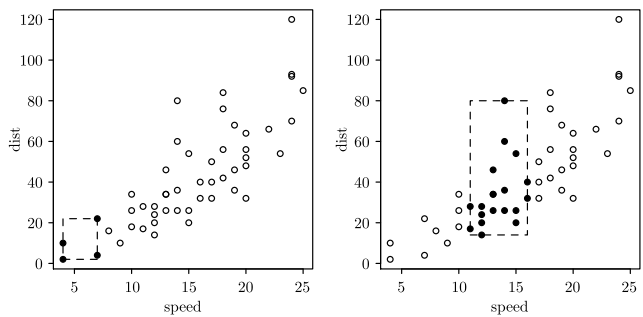}

%[Brush a scatterplot using the MVC design]
\caption{Brush a scatterplot using
the MVC design: brush the first four points (left), then brush all points
from the 10th to 28th. The dashed rectangle denotes the ``brush,''
which is
normally created by dragging the cursor over the points. The \texttt{cars}
data is a data set in base R. It has been ordered first by \texttt
{speed} and
then \texttt{dist} in the increasing order, so the bottom-left point
is the
first observation in the data.}\label{fig:mvc-brush}
\end{figure}

We bind a function \texttt{reactiveModel()} to the object \texttt
{reactiveBrush} through the base R function \texttt{makeActiveBinding()}.
When we assign new values to the object \texttt{reactiveBrush}, the function
defined in \texttt{reactiveModel()} will be called: inside the
function, the
logical variable \texttt{b} is modified by the indices \texttt{i} and the
view is updated accordingly. The two lines below achieve the same goal as
the MVC example in Figure~\ref{fig:mvc-brush}.

%f6 #&#
\begin{figure}[h]

\includegraphics{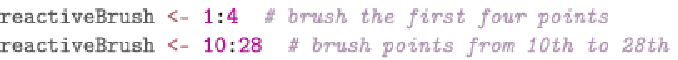}

  %\begin{knitrout}
%{fgcolor} \ifinner\ifhmode \def\at@end@of@kframe{\end{minipage}}
%%
%-\width
%%
%%
  %\caption{}
  \label{fig6}
\end{figure}\vspace*{-6pt}

%f7 #&#

Now our only task is to assign indices of the brushed points to \texttt
{reactiveBrush}, since the plot will be updated automatically. A real
interactive graphics system is more complicated than the above toy example,
but it shows the foundation of the pipeline. The two kinds of interactive
objects in \textbf{cranvas} are explained in the next two sections,
respectively.

%s3.1 #&#
\subsection{Mutaframes}

A mutaframe is an extension to the R data frame. They are mutable,
which means
that changes to its elements can be made anywhere regardless of the current
environment. By comparison, a data frame can only be modified in the
environment in which it was created, unless we use the nonlocal assignment
operator \texttt{<\textcompwordmark{}<-}. The difference is
highlighted in
the example below.

%f8 #&#
\begin{figure}[h]

\includegraphics{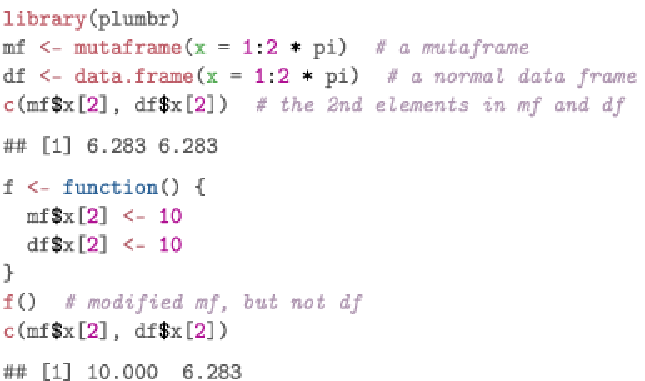}

%{fgcolor} \ifinner\ifhmode \def\at@end@of@kframe{\end{minipage}}
%%
%-\width
%%
%{=} \hlnum{1}\hlopt{:}\hlnum{2} \hlopt{*} \hlstd{pi)} \hlcom{# a
%mutaframe}
%{=} \hlnum{1}\hlopt{:}\hlnum{2} \hlopt{*} \hlstd{pi)} \hlcom{# a
%normal data frame}
%df}\hlopt{$}\hlstd{x[}\hlnum{2}\hlstd{])} \hlcom{# the 2nd
%elements in mf and df}
%%
%## [1] 6.283 6.283
%%
%df}\hlopt{$}\hlstd{x[}\hlnum{2}\hlstd{])}
%%
%## [1] 10.000 6.283
%%
  %\caption{}
  \label{fig8}
\end{figure}

As we can see, \texttt{mf} can be modified inside \texttt{f()}, but
\texttt
{df} cannot, therefore, we can share the same mutaframe across multiple
plots. Another important feature of mutaframes is that we can attach
listeners to them. A listener is essentially an R function which is called
upon changes in the mutaframe. For interactive graphics, views are updated
with listeners. Below
we create a mutaframe and attach a listener to it to
redraw the scatterplot.

%f9 #&#
\begin{figure}[h]

\includegraphics{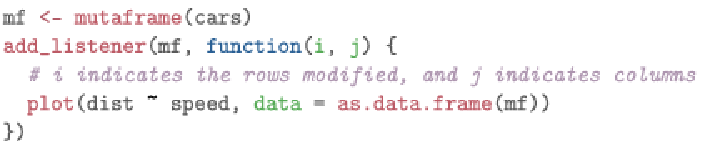}

%{fgcolor} \ifinner\ifhmode \def\at@end@of@kframe{\end{minipage}}
%%
%-\width
%%
%{i}\hlstd{,} \hlkwc{j}\hlstd{) \{}
%%
  %\caption{}
  \label{fig9}
\end{figure}

\begin{figure}

\includegraphics{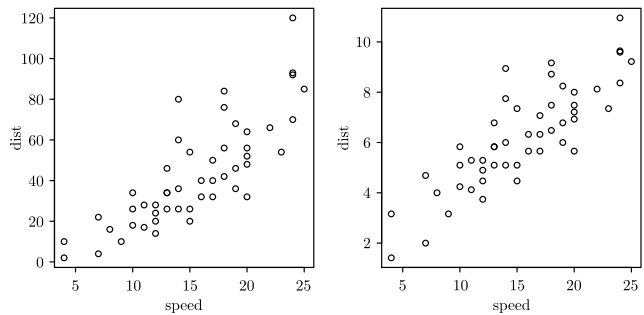}

%[The original scatterplot is automatically updated]
\caption{The original
scatterplot (left) is automatically updated (right) when the \texttt{dist}
variable is square-root transformed (right). We can also modify the
\texttt{speed} variable or change the values of some rows in the data to update
the plot.}\label{fig:mf-plots}%\vspace*{10pt}
\end{figure}

Now whenever we update \texttt{mf}, the scatterplot will be updated
accordingly. For example,
we make a square-root transformation of the
\texttt
{dist} variable (see Figure~\ref{fig:mf-plots} for the original plot and
the transformed version).

%f10 #&#
\begin{figure}[h]

\includegraphics{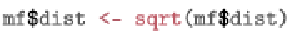}

%{fgcolor} \ifinner\ifhmode \def\at@end@of@kframe{\end{minipage}}
%%
%-\width
%%
%{(mf}\hlopt{$}\hlstd{dist)}
%%
  %\caption{}
  \label{fig10}\vspace*{-10pt}
\end{figure}

%f11 #&#

%f12 #&#
\begin{figure}[b]%\vspace*{10pt}

\includegraphics{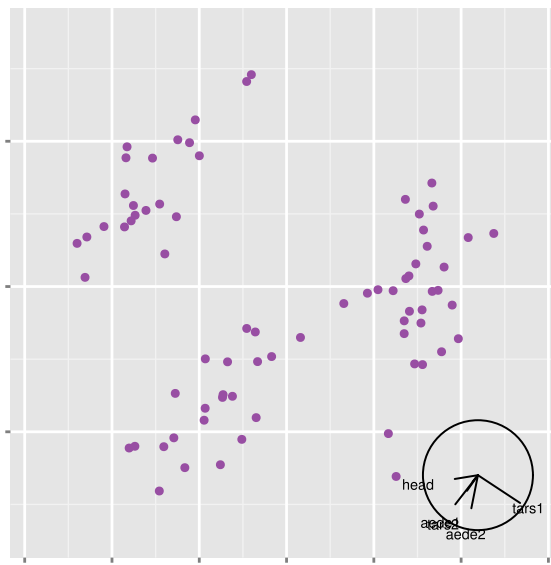}

%[A grand tour through the flea data]
\caption{A grand tour through the flea
data. The points are separated into three clusters. Watch the video online
at \protect\url{http://cranvas.org/examples/qtour.html}.}\label{tour}
\end{figure}

A more complex but direct application of mutaframes is the example
shown in
the movie displayed in Figure~\ref{tour}: here, we see a two-dimensional
grand tour \citep{asimov1985} through the flea data set provided in the
\textbf{tourr} package \citep{tourr}. A two-dimensional tour consists of
a series of projections into two-dimensional space. By choosing close
consecutive projections,
a sense of continuity is preserved for the observer. This
continuity allows us to identify clusters as groups of points that
share a
common fate (e.g., \cite{wolfe2012sensation}). Internally, the movie
is created by repeated changes to the $X$ and $Y$ values displayed in a
scatterplot,
which are propagated to the view.%\vadjust{\goodbreak}

%f13 #&#
\begin{figure}[h]

\includegraphics{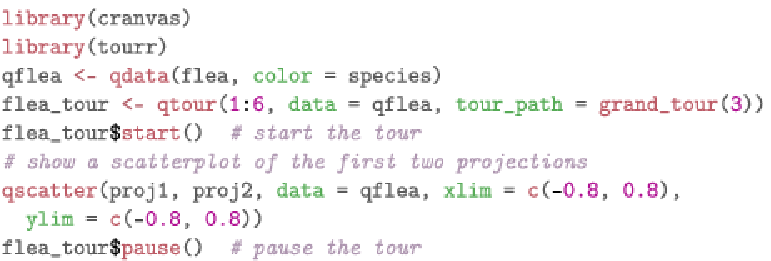}

%{fgcolor} \ifinner\ifhmode \def\at@end@of@kframe{\end{minipage}}
%%
%-\width
%%
%{:}\hlnum{6}\hlstd{,} \hlkwc{data} \hlstd{= qflea,} \hlkwc
%{tour_path} \hlstd{=} \hlkwd{grand_tour}\hlstd{(}\hlnum{3}\hlstd{))}
%the tour}
%qflea,} \hlkwc{xlim} \hlstd{=} \hlkwd{c}\hlstd{(}\hlopt{-}\hlnum
%{0.8}\hlstd{,} \hlnum{0.8}\hlstd{),}
%{0.8}\hlstd{,} \hlnum{0.8}\hlstd{))}
%the tour}
%%
  %\caption{}
  \label{fig13}
\end{figure}

Interactivity of a mutaframe can be propagated to its subsets, which allows
multiple applications based on one mutaframe and its offsprings. For
instance, we can select a subset of points in a scatterplot, obtain their
indices and use the indices to subset the original mutaframe to draw a new
plot. The new plot is then automatically connected with the original plot:
when we interact with the new plot, the selection will be passed to the
mutaframe and propagated to the original plot. This is similar to an example
described in early work by \citet{hurley1988}.

%s3.2 #&#
\subsection{Reference Classes}

R reference classes were introduced in R version 2.12. This made it
possible to
create objects with fields that can be accessed by reference. A consequence
of this feature is that such objects can be used for storing metadata in
the graphics system, and the data can be modified outside of plotting
functions. For instance, we can store the axis limits in an object
\texttt{meta} as \texttt{meta\$limits}. In the terminology of reference classes,
\texttt{limits} is called a \emph{field} of \texttt{meta}. After the plot
has been drawn, we are still able to modify its limits and the new limits
will be available to the internal drawing subroutines of the plotting
function. This is inconvenient, if not impossible, under the usual
copy-on-modify semantics in R. The brushing example in Figure~\ref{fig:mvc-brush}
is rewritten using reference classes.

%f14 #&#
\begin{figure}[h]

\includegraphics{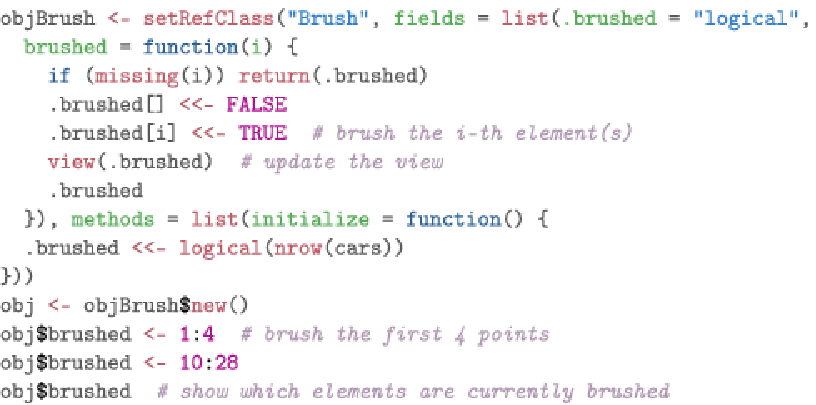}

%{fgcolor} \ifinner\ifhmode \def\at@end@of@kframe{\end{minipage}}
%%
%-\width
%%
%{"Brush"}\hlstd{,} \hlkwc{fields} \hlstd{=} \hlkwd{list}\hlstd
%{(}\hlkwc{.brushed} \hlstd{=} \hlstr{"logical"}\hlstd{,}
%{) \{}
%{return}\hlstd{(.brushed)}
%i-th element(s)}
%{initialize} \hlstd{=} \hlkwa{function}\hlstd{() \{}
%{nrow}\hlstd{(cars))}
%{:}\hlnum{4} \hlcom{# brush the first 4 points}
%{:}\hlnum{28}
%are currently brushed}
%%
  %\caption{}
  \label{fig14}
\end{figure}

We created a reference class object \texttt{obj} from the constructor
\texttt
{objBrush}, and this object has a field called \texttt{.brushed} which
is a
logical vector to store the brush status. The other field \texttt{brushed}
is a function that acts as the controller: we can assign new values to it,
and the view will be updated accordingly. We can also query the current
brush status to, for example, explore the brushed subset of the data
separately. The object \texttt{obj} can be modified anywhere in the system
as desired, which is often not the case for normal R objects. We will show
how reference classes work for interactions in single display applications
later.

What is more important is the extension by the \textbf{objectSignals}
package \citep{objectSignals} based on reference classes. The objects
created from this package are called ``signal objects,'' which are basically
special reference classes objects with listeners attached on them. This is
similar to mutaframes described before, but we can create objects of
arbitrary structures. The difference between mutaframes and signal objects
is similar to the difference between data frames and lists in R.

%s3.3 #&#
\subsection{Reactive Programming Behind Cranvas}\label{sec:reactive-cranvas}

Mutaframes and reference classes objects are extensively used in
\textbf{cranvas}, although this may not be immediately obvious
to the users. Below
we show some quick examples based on the Ames housing data
(Ames, IA, 2008--2012). Before we draw any plots in \textbf{cranvas}, we have to
create a
mutaframe using the function \texttt{qdata()}. %See Figure~\ref{fig15}.

%f15 #&#
\begin{figure}[h]

\includegraphics{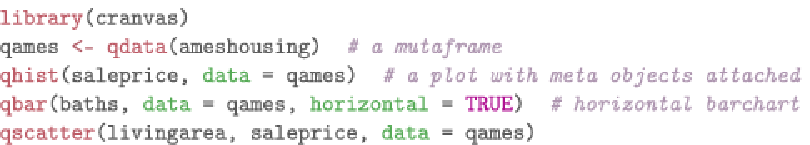}

%{fgcolor} \ifinner\ifhmode \def\at@end@of@kframe{\end{minipage}}
%%
%-\width
%%
%{# a mutaframe}
%{horizontal} \hlstd{=} \hlnum{TRUE}\hlstd{)} \hlcom{# horizontal barchart}
%{= qames)}
%%
  %\caption{}
  \label{fig15}
\end{figure}

%f16 #&#
\begin{figure}[b]

\includegraphics{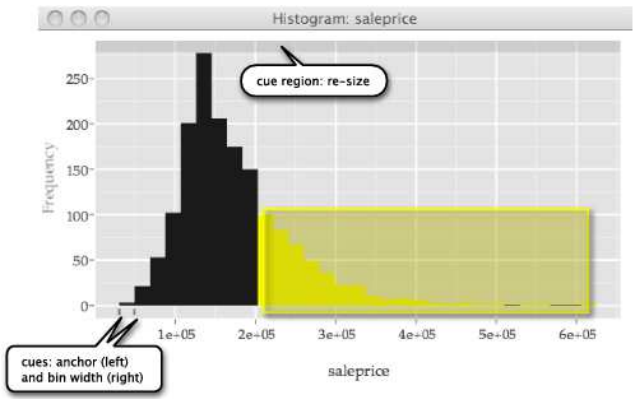}

%[Histogram of sales prices]
\caption{Histogram of sales prices. Sales
of \$200k and more are selected and highlighted in yellow. Markers show
visual cues.
See \protect\url{http://cranvas.org/examples/qhist.html} for a video of
the interactions.}\label{fig:hist}
\end{figure}

%f17 #&#
\begin{figure*}[b]

\includegraphics{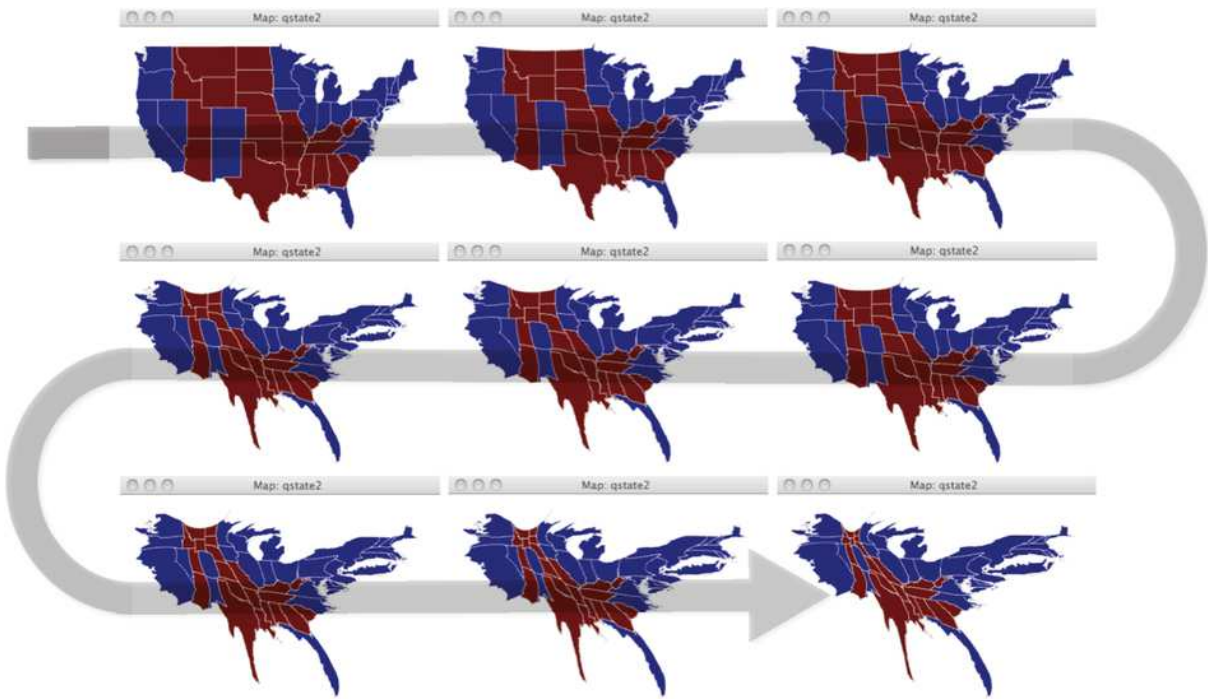}

%[Morph from a choropleth chart of the US to a population based
%cartogram]
\caption{Morph from a choropleth chart of the US (top left) to a
population-based cartogram (bottom right). The color represents electoral
votes of states toward the Democratic (blue) or Republican party (red) in
the 2012 Presidential election. The arrow indicates the direction of the
morph.}\label{fig:cartogram}%\vspace*{-20pt}
\end{figure*}

The function \texttt{qhist()} draws a histogram of the sale price,
\texttt
{qbar()} draws a bar chart of the number of bathrooms, and \texttt
{qscatter()} draws a scatterplot of the sale price against the living area.
All plotting functions have to take a \texttt{data} argument, which is a
mutaframe. Inside each function, listeners will be built on the data so
changes in the plot can be propagated back to the data object and further
passed to other plots.

The returned value of a plotting function contains the signal object, which
can be retrieved from the attributes of the returned value. The user can
manipulate the signal object and the plot can respond to the changes
because a number of listeners have been attached to it internally when we
call the plotting function.

Figures~\ref{fig:hist} and~\ref{fig:bar} show the histogram and the
bar chart
from the above R code. We will present details about the reactive
objects in
the next section.

%%==============================================================================%
%s4 #&#
\section{An Anatomy of Interactions}\label{sec:interaction}
%%==============================================================================%

In this section we show how some common interactions, including brushing,
zooming and querying, etc, were implemented in \textbf{cranvas}. The data
infrastructure is based on mutaframes and reference classes/signal objects,
as introduced in the previous section. The actual drawing is based on the
packages \textbf{qtbase} \citep{qtbase} and \textbf{qtpaint} \citep
{qtpaint}, which provide an interface from R to Qt \citep{verzani2012}.

%s4.1 #&#
\subsection{Input Actions}

Interaction with a system involves user actions as the input to the system,
then the system resolves the input information and responds to the user.
Interaction happens on multiple levels of user actions. The most common
forms of interaction with a display are listed below in decreasing
order of
immediacy with which this interaction between the user and the display
happens:

\textit{Direct manipulation} of graphical objects (Shneiderman, \citeyear{Shneiderman1983};
\cite{swayne1999}; \cite{wills1999}) is at the heart of interactive graphics. Direct
manipulation is what we use only for the highest level of interaction, such
as selection or brushing of elements. With a set of different modes
(querying, scaling mode as, for example, implemented in XGobi/GGobi), a set
of different or additional interactions can be incorporated at the highest
level. Another approach is to make use of visual cues, which suggest
available interactions to the user, for example, changing the cursor upon
entering the cue area. Visual cues are usually associated with changes to
the resolution of a representation or scales of a display. Figure~\ref{fig:hist}
shows an example of a plot with visual cues.

\textit{Input devices} such as a mouse or touch pad allow
interaction beyond click selection. Most toolkits
support wheel events (either through the presence of a mouse wheel, a mouse
move with an additional modifier key or a touch gesture), and a wheel event
often corresponds to the zooming of a plot.

\textit{Keystrokes} can be used as shortcuts and for quick access to functions.
Figure~\ref{fig:cartogram} shows an example of a key stroke interaction (arrow keys
\textsf{Left} and \textsf{Right}) to move between a choropleth map of the
United States and a population-based cartogram.

\textit{Functional access} through the command-line: Accessor functions
allow us
to get information about the state of objects (e.g., get the indices of
selected elements). Mutator functions enable the user to set a particular
state for objects in a display (e.g., set highlighting color to red, set
points to size 5). We call this level of interaction ``indirect
manipulation'' of graphics.

On the developer's side, the main idea behind resolving an interaction
between the user and the display is to actually resolve the interaction at
the level of the data, but make it appear as if the user had directly
interacted with the graphical object. This is essentially what happens in
\textbf{cranvas} when we interact with plots.

All levels of interaction above are supported in \textbf
{cranvas}, and both direct and indirect manipulation
are available. At its core, all kinds of manipulation end up as
changes to the underlying data objects, which is described in
the next section.

%f18 #&#
\begin{figure}%\vspace*{30pt}

\includegraphics{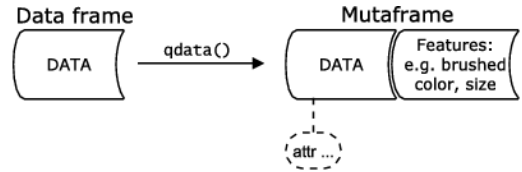}

%[The conversion from a data frame to a mutaframe]
\caption{The conversion
from a data frame to a mutaframe, which can be imagined as an augmented and
mutable data frame. Additional columns for the brush status and aesthetics
of graphical elements are appended to the original data frame.}\label{fig:mutaframe}%\vspace*{20pt}
\end{figure}

%s4.2 #&#
\subsection{Reactive Data Objects}\label{sec:reactive-data}

Figure~\ref{fig:mutaframe} illustrates the first step in \textbf{cranvas},
to create a mutaframe. The function
\texttt{qdata()} in \textbf{cranvas} returns a mutaframe with additional
columns. Below is a simple example.

%f19 #&#
\begin{figure}[h]

\includegraphics{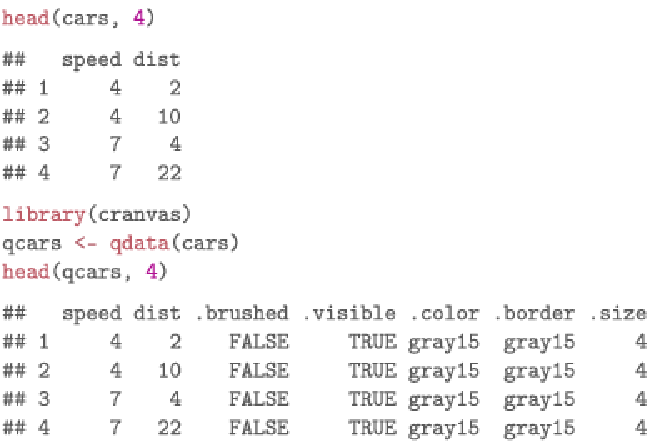}

%{fgcolor} \ifinner\ifhmode \def\at@end@of@kframe{\end{minipage}}
%%
%-\width
%%
%%
%## speed dist
%## 1 4 2
%## 2 4 10
%## 3 7 4
%## 4 7 22
%%
%%
%## speed dist .brushed .visible .color .border .size
%## 1 4 2 FALSE TRUE gray15 gray15 4
%## 2 4 10 FALSE TRUE gray15 gray15 4
%## 3 7 4 FALSE TRUE gray15 gray15 4
%## 4 7 22 FALSE TRUE gray15 gray15 4
%%
  %\caption{}
  \label{fig19}
\end{figure}

%f20 #&#
\begin{figure*}%[b]\vspace*{10pt}

\includegraphics{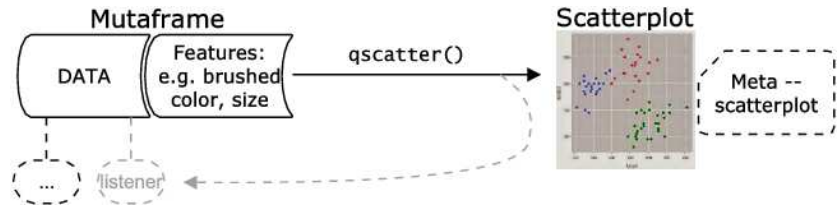}

%[Create a scatter plot and attach a meta object to it]
\caption{Create a
scatterplot and attach a metaobject to it.}\label{fig:meta}
\end{figure*}
The data frame \texttt{cars} was augmented by columns such as
\texttt{.brushed} and \texttt{.color}. The \texttt{.brushed} column
indicates the brush status of graphical elements (\texttt{TRUE} means an
element is brushed), and \texttt{.color} stores the colors of elements.
It is up to a specific plot how to interpret these additional
columns. For example, in scatterplots, because each row in the data
corresponds to a point in the plot, points can be colored by \texttt{.color}
and highlighted by the logical in \texttt{.brushed}. For a bar chart,
displaying frequencies of a categorical variable, \texttt
{.brushed} may result in a partially or fully highlighted bar when only a
subset in a category is brushed.

Each single display application in \textbf{cranvas} creates a plot and
attaches listeners on the mutaframe at the same time. Figure~\ref{fig:meta}
shows how a scatterplot is created from a mutaframe: before the \texttt
{qscatter()} function displays the plot, it binds the augmented columns in
the mutaframe with the plot layers using listener functions, so that when
these columns are updated, the plot can be updated. The Qt graphics
framework allows
us to build a plot using layers, which makes it possible to update one
component of a plot without having to update all others. This gives us
a lot
of performance gain, especially when we interact with plots of large numbers
of elements. In \textbf{cranvas}, the \texttt{.color} column should update
the main layer of points and the \texttt{.brushed} column controls the brush
layer. See \url{http://cranvas.org/2013/10/qt-performance/} for an
example of brushing a scatterplot of three million points, which takes less
than 0.01 second to render, and is highly responsive to the brush. The main
plot layer of three million points is not redrawn when the brush moves over
the plot, only the brush layer.

The other type of reactive data objects in \textbf{cranvas} are the metadata.
Such objects often contain plot-specific information, such as the
names of variables in the plot and the axis limits, etc. When
a plot is created, a copy of metadata is generated and associated with it.
Zooming (\url{http://cranvas.org/examples/qscatter.html}) is an example.
Behind the scenes the axis limits are modified in the
metadata based on the mouse wheel event.

Since the data structure of metaobjects is flexible, its application
can be broad. The \textbf{cranvas} package allows adding or customizing
meta information to any displays. For example, the user can specify a
function in the metaobject to generate text labels when querying a
plot. In
the following text, we use \texttt{meta} to denote a metadata object.

%s4.3 #&#
\subsection{Interactions}

This section describes the interactions supported in \textbf
{cranvas} and how they are related to the reactive data objects.

%s4.3.1 #&#
\subsubsection{Brushing and selection}

Brushing and selection are interactions that highlight a subset of graphical
elements in a plot. It is usually achieved by dragging a rectangle
(or other closed shape) over a plot and the elements inside the
rectangle are
selected. The rectangle, when in the brushing mode, is persistent on
the screen.
For the selection mode, the rectangle is transient, meaning that it
disappears when the mouse is released.

We use the \texttt{qscatter()} function to illustrate the basic idea. We
show a sketch using the pseudo code below.\vadjust{\goodbreak}

%f21 #&#
\begin{figure}[h]

\includegraphics{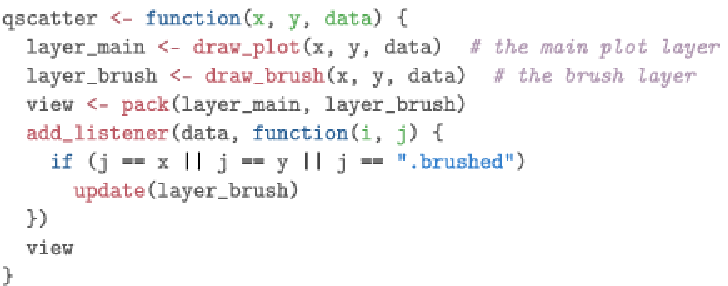}

%{fgcolor} \ifinner\ifhmode \def\at@end@of@kframe{\end{minipage}}
%%
%-\width
%%
%{x}\hlstd{,} \hlkwc{y}\hlstd{,} \hlkwc{data}\hlstd{) \{}
%data)} \hlcom{# the brush layer}
%{i}\hlstd{,} \hlkwc{j}\hlstd{) \{}
%{".brushed"}\hlstd{)}
%%
  %\caption{}
  \label{fig21}
\end{figure}

%f22 #&#
\begin{figure*}

\includegraphics{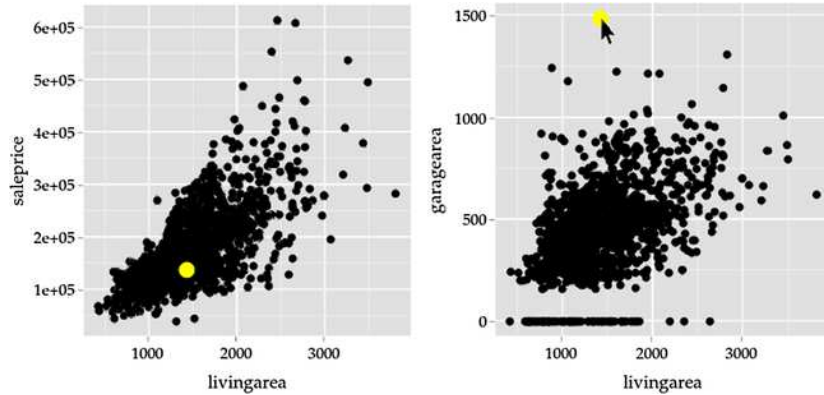}

%[One-to-one linking]
\caption{One-to-one linking: we highlight the
house with
the largest garage in the right plot and this house is also highlighted in
the left plot. Apparently it is not of the highest price.}\label{fig:one-one}
\end{figure*}

There are two layers \texttt{layer\_main} and \texttt{layer\_}\texttt{brush}
in the
plot. The brush layer is used to redraw the brushed points only, so
that the
main layer can stay untouched when points are highlighted. The key for
brushing/selection is the listener added to the mutaframe \texttt
{data} by
\texttt{add\_listener()}: when the column \texttt{x}, \texttt{y} or
\texttt
{.brushed} is modified, the brush layer is updated (changes in other columns
will not affect the plot). Adding the listener is denoted by the dashed
arrow in
Figure~\ref{fig:meta}.

When the plot is brushed, the points under the brush are identified by the
mouse events in Qt and the \texttt{.brushed} variable in the
mutaframe is modified. Because of the listener associated with \texttt
{.brushed},
the brush layer will be redrawn. Therefore, the selection is actually handled
with one step of backtracking: once the user draws a selection
rectangle, we
update \texttt{.brushed} immediately, which triggers the update of the brush
layer. Because this occurs in a fast pace, the user may have an illusion
that the cursor directly selected the points. See Figure~\ref{fig:one-one}
for an example of brushing scatterplots of the Ames housing data (R
code is
provided in the next section). The selection mode was used in these plots,
so we do not see the brush rectangle. Brushing mode was used in
Figure~\ref{fig:hist} and the yellow rectangle illustrates brush position.

In the case of a histogram, bins are the graphical objects intersecting with
a selection rectangle. The backtracking corresponds to identifying all
records in the mutaframe falling within the limits of the selected bin. The
binary variable \texttt{.brushed} is changed when the brush moves over the
bins, and the change is propagated to all dependencies, which results
in an
update to all dependent views. One of the dependent views is the histogram
itself, which shows highlighting in the form of superimposing a
histogram of the
highlighted records on top of the original histogram. What the user
perceives as ``selecting'' bins is actually a reaction to a change in the
internal brushing variable.

%s4.3.2 #&#
\subsubsection{Linking}

Linking forms the core of communication between multiple views. By default,
all views that involve variables from the same mutaframe are linked. Linking
within the same data set is implicitly one-to-one linking.

Using the mutaframe \texttt{qames} from
Section~\ref{sec:reactive-cranvas}, two scatterplots are created
(Figure~\ref{fig:one-one}) and illustrate one-to-one linking.

%f23 #&#
\begin{figure}[h]

\includegraphics{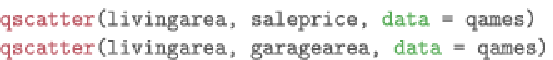}

%{fgcolor} \ifinner\ifhmode \def\at@end@of@kframe{\end{minipage}}
%%
%-\width
%%
%{= qames)}
%{= qames)}
%%
  %\caption{}
  \label{fig23}
\end{figure}

%f24 #&#
\begin{figure*}

\includegraphics{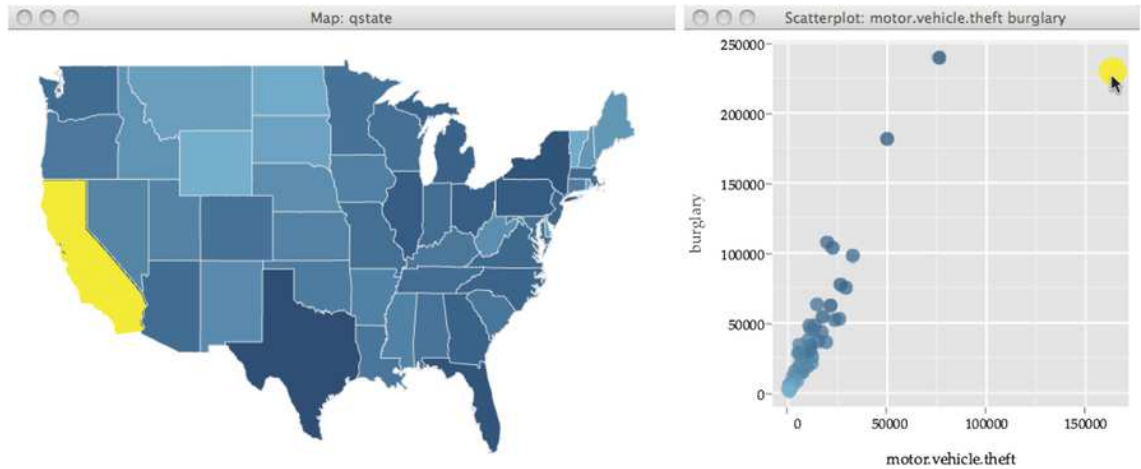}

%[Linked map and scatterplot]
\caption{Linked map (left) and scatterplot
(right). Color shading shows (log) state population with darker shades
indicating higher values. The scatterplot on the right displays the number
of burglaries in 2009 versus the number of motor vehicle thefts by state.
There is a strong correlation between the variables, which is mainly induced
by a strong underlying correlation with population. Compared to other
states, California displays a high number of motor vehicle thefts compared
to the number of burglaries.}\label{fig:linked}
\end{figure*}

The selected property has extraordinarily large ga\-rage area, for the living
area, and has a below average sale price.

Recall from the previous section that when a scatterplot is created, a
listener to update the brush layer is attached to the mutaframe. It
does not
matter where the mutaframe is modified, all the brush layers will be updated
if the \texttt{.brushed} variable in the mutaframe is modified. When we
interact with either of the plots, the other plot will respond to the
changes because both plots depend on the \texttt{.brushed} variable in the
same mutaframe.

It is feasible to extend this concept to link different sources or
aggregation levels of the data. Take the following two types of
linking, for
example:

\textit{Categorical linking} means when we brush one or more
observations in
one category, all observations in this category are brushed; this is
achieved in the listener by setting all elements of \texttt{.brushed} in
this category to \texttt{TRUE};

\textit{kNN linking} ($k$ nearest neighbor) means when we brush an
observation, its $k$ nearest neighbors under a certain distance metric are
brushed as well; again, this is nothing but setting the relevant elements
in \texttt{.brushed} to \texttt{TRUE}.\vadjust{\goodbreak}

They can be applied to a single data source (called ``self-linking'') or
multiple data sources. In the latter case, the listener in one data object
needs to update other data objects. In Figure~\ref{fig:linked}, the
map and
the scatterplot use two different sources, and they are linked via
categorical linking through the state names. Each state in the map is
described by multiple points defining the state boundary, but each
state has
only one observation in the scatterplot. If we brush California in the
scatterplot, the whole polygon of California (containing multiple locations)
is highlighted. On the other hand, if we brush a part of a state in the map,
that means the whole state should be highlighted, which is an example of
self-linking.

\begin{figure*}

\includegraphics{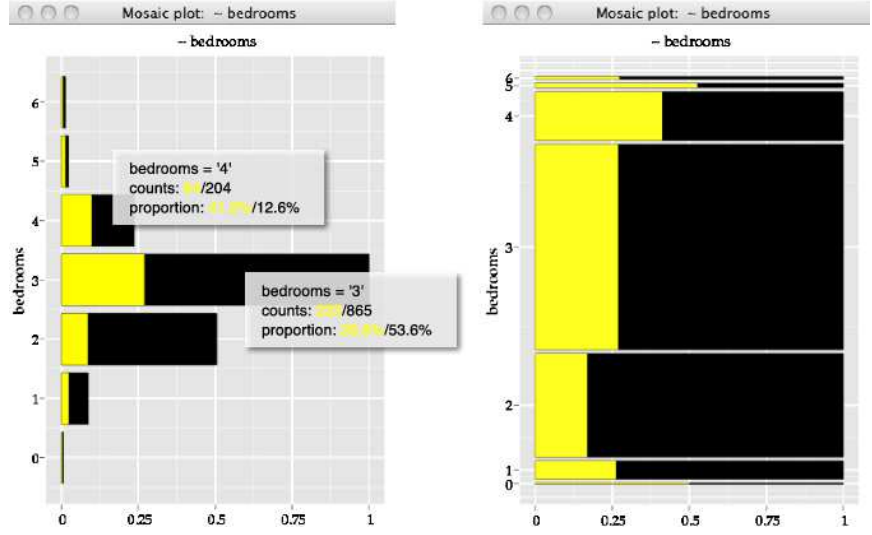}

%[Bar chart and spine plot of the number of bedrooms in housing sales]
\caption{Bar chart (left) and spine plot (right) of the number of bedrooms in
housing sales. Highlighted are sales of \$200k and higher. Querying gives
details on each bin (houses with a particular number of bedrooms) and the
selected houses with this bin.}\label{fig:bar}
\end{figure*}

Linking can also be done on the same data with different aggregation levels,
such as the raw data and binned data. The histogram in Figure~\ref{fig:hist}
shows sales prices of all houses sold in the Ames housing data. The yellow
rectangle corresponds to an area-based selection of all houses with
sales of
\$200k or more, which triggers a highlighting of corresponding houses in
all displays of the Ames housing data. This includes the histogram itself,
where all bins intersecting with the selection rectangle are filled with
highlighting color, and all bar charts of the number of bedrooms as
shown in
Figure~\ref{fig:bar}.

%f25 #&#

%s4.3.3 #&#
\subsubsection{Zooming and panning}

Zooming and panning change or shift the scale of the view, so we can see the
data at different resolutions. This can be resolved directly without
the need of interaction with the original data. In \textbf{cranvas}, the
core of zooming and panning consists of simple changes to the metadata
\texttt{meta\$limits}. This is illustrated with the following pseudo
code.\vadjust{\goodbreak}

%f26 #&#
\begin{figure}[h]

\includegraphics{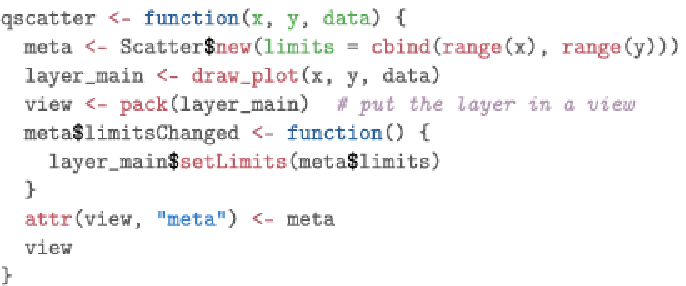}

%{fgcolor} \ifinner\ifhmode \def\at@end@of@kframe{\end{minipage}}
%%
%-\width
%%
%{x}\hlstd{,} \hlkwc{y}\hlstd{,} \hlkwc{data}\hlstd{) \{}
%{(}\hlkwc{limits} \hlstd{=} \hlkwd{cbind}\hlstd{(}\hlkwd
%{range}\hlstd{(x),} \hlkwd{range}\hlstd{(y)))}
%put the layer in a view}
%{function}\hlstd{() \{}
%{$}\hlstd{limits)}
%%
  %\caption{}
  \label{fig26}
\end{figure}

\texttt{Scatter} is a constructor created from reference classes, containing
a field named \texttt{limits} in \texttt{meta}. The key here is to
set up
the event \texttt{meta\$limitsChanged}: this event is triggered when
\texttt
{meta\$limits} is modified and the axis limits of the main plot layer are
replaced by the new value of \texttt{meta\$limits}. The \texttt{setLimits()}
method is from Qt, which is used to set new limits on a layer, and Qt will
update the view when the limits are changed.

%f27 #&#
\begin{figure*}[b]

\includegraphics{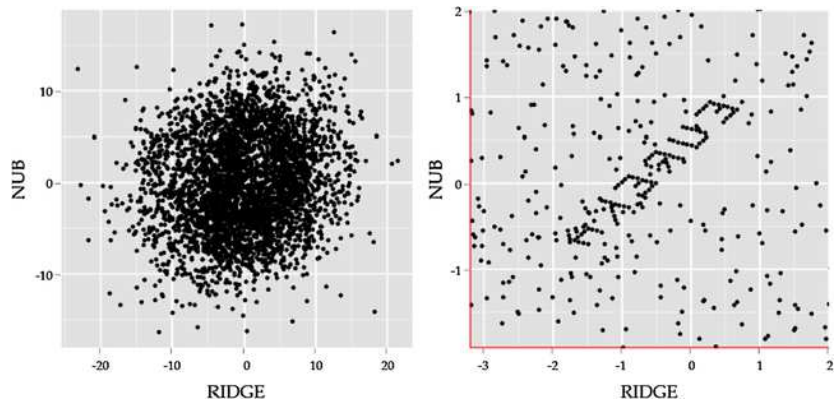}

%[Zooming into the \texttt{pollen} data]
\caption{Zooming into the
\texttt
{pollen} data. The original scatterplot (left) does not show anything
interesting, but as we zoom into the plot, we see a hidden word ``EUREKA.''}\label{fig:zoom}
\end{figure*}

In \textbf{cranvas}, \texttt{meta\$limits} is modified by the mouse
wheel event
for zooming and by the mouse drag event for panning.
Figure~\ref{fig:zoom} shows two screenshots of the
original scatterplot and the zoomed version, respectively.

%f28 #&#
\begin{figure}[h]

\includegraphics{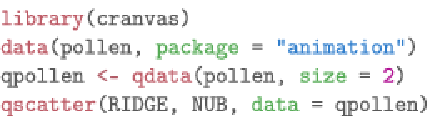}

%{fgcolor} \ifinner\ifhmode \def\at@end@of@kframe{\end{minipage}}
%%
%-\width
%%
%{"animation"}\hlstd{)}
%{size} \hlstd{=} \hlnum{2}\hlstd{)}
%%
  %\caption{}
  \label{fig28}
\end{figure}

A benefit of controlling the limits in this manner is that it is a local
property of the plot, enabling the user to examine different resolutions
in two different displays.

%s4.3.4 #&#
\subsubsection{Querying/identifying}

Querying/identifying is another interaction that can usually be resolved without
information from the original data.
Querying of graphical objects involves, in a first step, the
display of the corresponding values of the metadata. For the
bar charts in Figure~\ref{fig:bar}, querying of the bins displays
information, about the bin's level and the number of records it
encompasses, as
well as the proportion of the whole data that this bin contains.

%s4.3.5 #&#
\subsubsection{Visual cues}

Visual cues aid the user to learn about available interactions.
Figure~\ref{fig:hist} shows several examples of visual cues in a
histogram. Both the anchor point and bin width are graphical
representations of
plotting parameters for the histogram. The anchor of a histogram is the
lower limit of the leftmost bin. The bin width defines the interval at
which breaks are made. Interacting with either anchor or
bin width cues produces horizontal shifts, which reset the actual values
parametrizing the histogram. Changes to the anchor allow for testing of
instabilities in the display due to discreteness in the data. Bin-width
changes show the data at different levels of smoothness and therefore allow
for a visualization of ``big'' picture marginal distributions at large bin
widths and the investigation of small pattern features, such as multiple
modes and gaps in the data, at small bin widths. Examples for both of these
interactions are available as movies in the supplementary material of
this paper.

The visual cues in this case also correspond to meta elements. Specifically,
the bin limits are stored in \texttt{meta\$breaks} and the histogram layer
is connected to the \texttt{meta\$breaksChanged} event. The anchor modifies
\texttt{meta\$breaks} when we drag it.

%s5 #&#
\section{Conclusions}

The concept of MVC is made implicit in reactive programming. Reactive data
objects are used to manage the multiple views and interactions in the
\textbf{cranvas} package.

The \textbf{cranvas} package is built on graphics layers in Qt (frontend)
and reactive data objects in R (backend). The plotting pipeline is expressed
and attached to mutaframes as well as metadata objects. Using the reactive
programming model, the user does not\vadjust{\goodbreak} need to pay attention to the whole
pipeline, which makes it easy to extend this system. For example,
implementing the tour is simply redrawing a scatterplot of the projected
variables that keep changing because the mutaframe will update
the view automatically upon changes.

The future work of \textbf{cranvas} involves including more plot types
such as
hexbin plots and scatterplot matrices, allowing users to define how the
system reacts to changes and adding a GUI.
The \textbf{qtbase} package has made it easy to build Qt GUI's in R. The
GUI widgets can be connected to the plots via reactive data objects.
They do not need to know the internal structure of plots. This kind of
modularity will make the system easier to maintain and extend than past
graphics software.

% what reactive programming enables cranvas to do and other systems
%cannot
% easily do is interaction from command line besides direct manipulation


\section*{Supplementary Materials}

The \textbf{cranvas} package is available on GitHub at
\url{https://github.com/ggobi/cranvas}. We have movies
showing some interactions
that are available on the website \url{http://cranvas.org}.

%suskaldyti doi

% imsref loaded by audrone.aklyte, 2014-05-09 09:55:46
% imsref loaded by audrone.aklyte, 2014-05-09 10:21:01
%

\end{document}